\documentclass[8.5pt,twoside,twocolumn]{article}
\usepackage{cite}
\usepackage{mhchem}
\usepackage{times,mathptmx}
\usepackage{secsty}
\usepackage{balance} 

\usepackage{amsmath,graphicx} 
\usepackage{lastpage}
\usepackage[format=plain,singlelinecheck=false,font=small,labelfont=bf,labelsep=space]{caption} 

\usepackage{fancyhdr}
\usepackage{graphicx}
\usepackage{etoolbox,hyperref}

\providecommand{\keywords}[1]{\textbf{Keywords } #1}
\providecommand{\pacsnumbers}[1]{\textbf{PACS numbers } #1}
\begin{document}

\thispagestyle{plain}
\fancypagestyle{plain}{
\renewcommand{\headrulewidth}{1pt}}
\renewcommand{\thefootnote}{\fnsymbol{footnote}}
\renewcommand\footnoterule{\vspace*{1pt}%
\hrule width 3.4in height 0.4pt \vspace*{5pt}} 
\setcounter{secnumdepth}{5}

\makeatletter 
\def\subsubsection{\@startsection{subsubsection}{3}{10pt}{-1.25ex plus -1ex minus -.1ex}{0ex plus 0ex}{\normalsize\bf}} 
\def\paragraph{\@startsection{paragraph}{4}{10pt}{-1.25ex plus -1ex minus -.1ex}{0ex plus 0ex}{\normalsize\textit}} 
\renewcommand\@biblabel[1]{#1}            
\renewcommand\@makefntext[1]%
{\noindent\makebox[0pt][r]{\@thefnmark\,}#1}
\makeatother 
\renewcommand{\figurename}{\small{Fig.}~}
\sectionfont{\large}
\subsectionfont{\normalsize} 

\fancyfoot{}
\fancyfoot[RO]{\footnotesize{\sffamily{\thepage}}}
\fancyfoot[LE]{\footnotesize{\sffamily{\thepage}}}
\fancyhead{}
\renewcommand{\headrulewidth}{1pt} 
\renewcommand{\footrulewidth}{1pt}
\setlength{\arrayrulewidth}{1pt}
\setlength{\columnsep}{6.5mm}

\twocolumn[
  \begin{@twocolumnfalse}
\noindent\LARGE{\textbf{Hydrogen mean force and anharmonicity in polycrystalline and amorphous ice}}
\vspace{0.6cm}

\noindent\large{\textbf{A. Parmentier\textit{$^{a,*}$}, C. Andreani\textit{$^{a,b}$}, G. Romanelli\textit{$^{c,a}$}, J. J. Shephard\textit{$^{d,e}$}, C. G. Salzmann$^{d}$, and R. Senesi\textit{$^{a,b,\dag}$}}}\vspace{0.5cm}

\noindent \normalsize{\textbf{Abstract.} The hydrogen mean force from experimental neutron Compton profiles is derived using deep inelastic neutron scattering on amorphous and polycrystalline 
ice. The formalism of mean force is extended to probe its sensitivity to anharmonicity in the hydrogen-nucleus effective potential. The shape of the mean
force for amorphous and polycrystalline ice is primarily determined by the anisotropy of the underlying quasi-harmonic effective potential. The data 
from amorphous ice show an additional curvature reflecting the more pronounced anharmonicity of the effective potential with respect to that of ice I$\textit{h}$.
}

\vspace{0.5cm}


\vspace{0.2cm}

\pacsnumbers{61.05.F-, 61.05.fg}

\vspace{0.5cm}
 \end{@twocolumnfalse}
  ]

\footnotetext{\textit{$^{a}$~Universit\`{a} degli Studi di Roma Tor Vergata, Dip. di Fisica e Centro NAST, Via della Ricerca Scientifica 1, 00133 Roma, Italy}}
\footnotetext{\textit{$^{b}$~CNR-IPCF Sezione di Messina, v.le F. Stagno D'Alcontres 37, 98158 Messina, Italy}}
\footnotetext{\textit{$^{c}$~ISIS Facility, Rutherford Appleton Laboratory, Chilton, Didcot, Oxfordshire, OX11 0QX, UK}}
\footnotetext{\textit{$^{d}$~University College London, Dept. of Chemistry, 20 Gordon Street, London WC1H 0AJ, UK}}
\footnotetext{\textit{$^{e}$~Dept. of Chemistry, Durham University, South Road, Durham DH1 3LE, UK}}
\footnotetext{\textit{Corresponding author: $^{*}$~alexandra.parmentier@uniroma2.it}}
\footnotetext{\textit{Corresponding author: $^{\dag}$~roberto.senesi@uniroma2.it}}

\section{Introduction}

Nuclear quantum effects, such as the zero-point energy and its interplay with the anharmonic character of the hydrogen bond (HB), affect a large number of water's properties ranging from its microscopic structure and dynamics to its thermodynamic and chemical behavior \cite{Ceriotti2016}.

Recently, new experimental and simulation techniques have been used to probe the quantum state of hydrogen nuclei in water and water systems by examining the hydrogen nuclear momentum distribution, $n$($p$), and the hydrogen nuclear mean kinetic energy, $\langle E_K\rangle$. These physical quantities are influenced by quantum effects and can be uniquely accessed via high energy neutron scattering using the deep inelastic neutron scattering (DINS) technique \cite{Gunn,Andreani}. The DINS refers to a specific regime of inelastic neutron scattering in which the incident neutron energy is well above the binding energies of the scattering atoms. This condition is experimentally achieved at high energy $\hbar \omega$ ($\ge 1$ eV) and momentum $\hbar q$ transfers ($\ge 25$ $\textup{\AA}^{-1}$). It is a specific regime where the neutron-scattering process is theoretically described within the framework of the impulse approximation (IA) \cite{West,Andreani}, which is exact in the limit of infinite momentum transfer, $\hbar q$ \cite{Reiter_1985, Watson_1996}.

There are several reports on DINS experiments and theoretical studies of $n$($p$) lineshapes and $\langle E_K\rangle$ values of hydrogen and light nuclei in water systems and in a variety of other materials. In particular, $n$($p$) and $\langle E_K\rangle$ observables are routinely used to fingerprint changes in the hydrogen bond network in water and water systems. Most recent reviews on the experimental studies and the use of eV neutron spectroscopy to investigate the properties of light nuclei in water and complex materials can be found in Ref. \cite{Andreani} and Ref. \cite{Felix2016}, respectively. 
DINS measurements of these observables are benchmarked with the results of electronic density functionals used in path integral molecular dynamics (PIMD) for the description of hydrogen bonded systems in {\it ab initio} numerical simulations \cite{Morrone2008,Car1985,Marx1996,Lin2010,Ceriotti2016b}. The most recent examples are DINS and PIMD studies in ice and water \cite{Andreani2013,Senesi2013}, ice \cite{Flammini2012,Senesi2013}, supercritical water \cite{Pantalei2008,Andreani2013}, and supercooled water \cite{Andreani2016}. In these cases, simulations and DINS experimental results provide new information on the three-dimensional effective potential energy surface experienced by the hydrogen nucleus.
In this context, a relevant parameter is the mean force (MF) function, $f(x)$, which provides an insight into the forces in a molecular system and is used to describe the average force acting on an atomic particle by keeping all other particles in the system fixed. 
Indeed, as pointed out by Feynman \cite{Feynm1939}, many problems of the molecular structure are essentially concerned with forces, such as the stiffness of chemical bonds and geometrical arrangements due to repulsions and attractions between atoms.

The mean force is expressed in terms of the spherical end-to-end distribution, $\tilde{n}(x)$, i.e., the Fourier transform of $n(p)$ \cite{Lin2010,Flammini2012}. For example, in the DINS and simulation study on hexagonal ice in Ref. \cite{Flammini2012}, it has been reported how the $f$($x$) function can be derived from the experimental data and how the accuracy required to unambiguously resolve and extract the effective hydrogen nuclear potential can be evaluated. It is also shown how the derivation depends on the signal-to-noise ratio in the DINS count rate. The latter is a consequence of data uncertainties and error propagation derived from the sequence of experimental correction routines and data analysis procedures.

In this study, we extend the formalism of mean force as a direct, model-independent, non-parametric approach to probe the experimental sensitivity to anharmonicity in the hydrogen nuclear effective potential, in order to separate the effects of anharmonicity from those of molecular anisotropy. This is applied first to synthetic, spherically averaged, momentum distribution data from model systems representing the local hydrogen environment in harmonic anisotropic potentials, and then, as a first approximation, by adding anharmonicity along the bond direction by using a simple Morse potential. Finally, the formalism is applied to experimental data on amorphous and polycrystalline ices, showing that the local environment of the hydrogen nucleus in amorphous ices is characterized by a more pronounced anharmonicity of the hydrogen nuclear effective potential, in comparison to that in ice I$\textit{h}$.

\section{Response Function from DINS experiments}

The IA assumes that the scatterers recoil freely from the collision with neutrons, the inter-particle interaction in the final state being negligible. The regime can be regarded as a special case of the incoherent approximation, where, in the case
of high-energy collisions, a short-time expansion ($t \rightarrow 0$) of the atomic 
position operator, $\mathbf{R}(t)$, is applied to the position operator of any scatterer of 
mass $M$ and momentum $\mathbf{p}$, i.e., $\mathbf{R}(t) = \mathbf{R}(0) + \frac{t}{M}\mathbf{p}$ 
\cite{Andreani}. By applying the momentum- and energy-conservation laws, it can be shown that the energy
distribution of the scattered neutrons is directly related to the distribution
of particle momenta parallel to the wave vector transfer $\textit{q}$, and the resulting (incoherent) dynamic structure factor yields:
\begin{equation}
S(\mathbf{q},\omega)=\hbar\int{n(\mathbf{p})\delta\left(\hbar\omega-\hbar\omega_{r}-\frac{\hbar\mathbf{q\cdot p}}{M}\right)d\mathbf{p}} 
\label{IA}
\end{equation}

where $\hbar\omega$ is the energy transfer, and $\hbar\omega_{r}=\frac{\hbar^2q^2}{2M}$ is the recoil energy.

Using the West scaling formalism, the two dynamic variables $\omega$ and $\mathbf{q}$ can be coupled by introducing the West variable $y=\frac{1}{\hbar}\mathbf{p \cdot \hat{q}}=\frac{M}{\hbar^{2}q}(\hbar\omega-\hbar\omega_{r})$ 
\cite{Andreani}, so that Eq. \ref{IA} can be re-written as

\begin{equation}
S(\mathbf{q}, \omega)=\frac{M}{\hbar q}J(y,\mathbf{\hat{q}}), 
\end{equation}

where $J(y,\hat{\mathbf{q}})$ is the response function, or Neutron Compton Profile (NCP), within the IA framework \cite{Andreani,Gunn}:

\begin{equation}
J(y,\mathbf{\hat{q}})=\hbar\int{n(\mathbf{p})\delta\left(\hbar y-\mathbf{ \hat{q} \cdot p}\right)d\mathbf{p}}.
\label{Radon}
\end{equation}

$J(y,\mathbf{\hat{q}})$ represents the probability that the atomic nucleus has a momentum parallel to $\mathbf{\hat{q}}$ of magnitude between $\hbar y$ and $\hbar(y+dy)$.

For isotropic samples, the momentum distribution only depends on ${\left|\bf p\right|}$, 
and the $\hat{\bf q}$ direction becomes immaterial. Thus, the NCP is expressed by
$2 \pi \hbar\int_{|\hbar y|}^{\infty}{pn(p)dp}$, and the expression for $n(p)$ yields:

\begin{equation}
n(p)=-\frac{1}{2\pi\hbar^{3}y}\left[\frac{dJ(y)}{dy}\right]_{\hbar y=p}.
\end{equation}

The IA is exact only in the limit of infinite wave vector transfer. At finite values
of $\mathbf{q}$, deviations occur, which are caused by the localization of the 
scatterer in its final state due to surrounding atoms, and are termed final state effects (FSEs). 
This causes a broadening of $J(y)$ that resembles an instrumental-resolution effect \cite{Senesi}.
In the presence of FSEs, the $J(y)$ function shows an additional dependence on $q$,
which in the isotropic case is expressed as a $\frac{1}{q}$ power series \cite{Sears,FSE}:

\begin{equation}
J(y,q)=J(y)- \frac{A_3}{q} \frac{d^3J(y)}{dy^3}+....
\end{equation}

Currently, DINS measurements are carried out at the VESUVIO beamline at the ISIS pulsed neutron and muon source (Rutherford Appleton Laboratory, Chilton, Didcot, UK) using the time of flight technique \cite{Vesuvio}. VESUVIO is the only instrument designed to exploit the DINS technique at high energy- and momentum transfers.

A DINS experiment on a condensed-water sample yields an experimental NCP, $F_l (y,q)$, from each {\it l-th} detector, for the hydrogen (or oxygen) nuclei. For each detector element $l$, the experimental NCP, due to finite $q$ values in the neutron scattering process, retains the $q$ dependence and is related to the DINS count rate via the expression: 
\begin{equation} \label{fdy}
F_l(y,q)=\frac{BM}{E_0 \ I(E_0)} q\,C_l(t)
\end{equation}
where $E_0$ is the initial energy of the neutron, $I(E_0)$ is the incident neutron flux at energy $E_0$, and $B$ is a constant determined by taking into account several contributions: the detector solid angle, its efficiency at the final energy $E = E_1$, the time-energy Jacobian, the free-atom neutron cross section, and the number of particles hit by the neutron beam. DINS data sets of all samples are $y$-scaled according to Eq. (\ref{fdy}). 

In a DINS experiment the asymptotic IA profile, strictly valid in the limit of infinite $q$ (asymptotic regime),
is broadened for each individual {\it l-th} detector by finite $q$ corrections terms, $\Delta J_l(y, q)$, known as final
state effects (FSEs) and by the instrumental resolution function, $R_l(y,q)$ :
\begin{equation}
F_l(y,q)= [J (y) + \Delta J_l(y, q)] \otimes R_l(y,q).
\label{defDY}
\end{equation}
where $R_l(y,q)$ is determined using standard Monte Carlo routines available on VESUVIO. This equation is used to describe the experimental NCP of Eq.(\ref{fdy}) for each individual {\it l-th} detector, $F_l(y,q)$. Full details on DINS formalism, description of operation of the VESUVIO instrument and experimental set up, experimental corrections and data analysis are reported in References \cite{Andreani2016,Flammini2012,Vesuvio2}. 

\section{Potential of Mean Force}
\label{PMF}

The interparticle potential in a condensed system can be expressed as a sum of pairwise terms which depend on the relative coordinates between particles. A coordinate $R(\mathbf{\tilde{q}})$ is used to indicate a hydrogen bond, a torsional angle, or linear combinations of similar quantities \cite{Harmandaris}; $\mathbf{\tilde{q}}$ is the generalized vector coordinate, along which the free-energy profile can be determined.

The free-energy profile, referred to as the Potential of Mean Force (PMF), is defined as the potential energy arising from the average force acting
between two fixed particles, with the average taken over the ensemble of configurational states for the remaining $N-2$ particles.

Through the use of $R(\mathbf{\tilde{q}})$, one can define the system in a hypersurface within the phase space, allowing one to derive the free energy, $F_R(R')$, 
the partition function, $Z_R(R')$, and the end-to-end reaction-coordinate distribution function, $P_R(R')$. These functions yield \cite{Trzesniak}:

\begin{equation}
\mathit{Z}_R(R')=\frac{1}{h^{3N}N!} \int \int{e^{- \frac{\mathcal{H}(\mathbf{\tilde{p}},\mathbf{\tilde{q}})}{k_BT}}\delta(R'-R(\mathbf{\tilde{q}}))d \mathbf{\tilde{p}}d \mathbf{\tilde{q}}};
\end{equation}

\begin{equation}
P_R(R')=\frac{\mathit{Z}_R(R')}{\mathit{Z}}=\frac{\int \int{e^{- \frac{\mathcal{H}(\mathbf{\tilde{p}},\mathbf{q})}{k_BT}}\delta(R'-R(\mathbf{\tilde{q}}))d \mathbf{\tilde{p}}d \mathbf{\tilde{q}}}}{\int \int{e^{- \frac{\mathcal{H}(\mathbf{\tilde{p}},\mathbf{\tilde{q}})}{k_BT}}d \mathbf{\tilde{p}}d \mathbf{\tilde{q}}}};
\end{equation}

\begin{equation}
F_R(R')=-k_BTlnP_R(R')-k_BTln \mathit{Z},
\end{equation}

with $\mathit{f}$ being the mean force, related to the free energy by $\mathit{f}=- \frac{dF_R(R')}{dR'}$.

Following Lin {\it et al.} \cite{Lin2010}, one can express both the partition function and the momentum distribution, $n(p)$, in terms of the one body density matrix
$\rho(\mathbf{r}, \mathbf{r'})=\left\langle \mathbf{r|e^{-\beta \mathcal{H} }| \mathbf{r'}}  \right\rangle$:

\begin{equation}
\mathit{Z}=\int{d \mathbf{r}}\rho(\mathbf{r},\mathbf{r'})
\end{equation}

and 

\vspace{4mm}

${n(\mathbf{p})=}\frac{1}{(2 \pi \hbar)^3 \mathit{Z}} \int{d \mathbf{r} d \mathbf{r'}e^{\frac{i}{\hbar}\mathbf{p \cdot}(\mathbf{r}- \mathbf{r'}) }\rho(\mathbf{r},\mathbf{r'})}=$

\begin{equation}
=\frac{1}{(2 \pi \hbar)^3} \int{d \mathbf{x} e^{\frac{i}{\hbar}\mathbf{p \cdot}\mathbf{x} } \tilde{n}(\mathbf{x})}
\end{equation}

with

\begin{equation}
\tilde{n}(\mathbf{x})=\frac{1}{ \mathit{Z}} \int{d \mathbf{r} d \mathbf{r'} \delta (\mathbf{r}- \mathbf{r'}- \mathbf{x}) \rho(\mathbf{r},\mathbf{r'})}.
\end{equation}


A viable computational strategy in the investigation of a condensed system is the statistical sampling using the Feynman path integral (PI) representation: $\tilde{n}(\mathbf{x})$ is the end-to-end distribution derived by a sum over open paths whereas closed paths determine $\mathit{Z_R(R')}$ \cite{Ceperley}. 

In such a representation, the density matrix is expressed by

\begin{equation}
\rho(\mathbf{r}, \mathbf{r'})= \int_{\mathbf{r}(0)=\mathbf{r},\mathbf{r}(\beta \hbar)=\mathbf{r'}}{\mathcal{D}\mathbf{r}(\tau)e^{-\frac{1}{\hbar} \int_0^{\beta \hbar}d \tau \left( \frac{m\mathbf{\dot{r}}^2(\tau)}{2} + V[\mathbf{r}(\tau)] \right)}}
\end{equation}

with $\beta=\frac{1}{k_BT}$.

If the linear transformation $\mathbf{r}(\tau)=\mathbf{\tilde{r}}(\tau)+y(\tau)\mathbf{x}$ is carried out
in the path space, then one can express the NCP in terms of the distribution $\tilde{n}(\mathbf{x})$. This action reshapes the open path $\mathbf{r}(\tau)$ into the closed path $\mathbf{\tilde{r}}(\tau)$, with the free particle contribution coming from the derivative of $y(\tau)$.
Thus, the end-to-end distribution is given by:

$\tilde{n}(\mathbf{x})=\frac{\int_{\mathbf{r}(0)-\mathbf{r}(\beta \hbar)=\mathbf{x}}{\mathcal{D}\mathbf{r}(\tau)e^{-\frac{1}{\hbar} \int_0^{\beta \hbar}d \tau \left( \frac{m\mathbf{\dot{r}}^2(\tau)}{2} + V[\mathbf{r}(\tau)] \right)}}}{\int_{\mathbf{r}(\beta \hbar)=\mathbf{r}(0)}{\mathcal{D}\mathbf{r}(\tau)e^{-\frac{1}{\hbar} \int_0^{\beta \hbar}d \tau \left( \frac{m\mathbf{\dot{r}}^2(\tau)}{2} + V[\mathbf{r}(\tau)] \right)}}}=$

\begin{equation}
=e^{-\frac{m\mathbf{x}^2}{2\beta \hbar^2}} \frac{\int_{\tilde{\mathbf{r}}(\beta \hbar)=\tilde{\mathbf{r}}(0)}{\mathcal{D}\tilde{\mathbf{r}}(\tau)e^{-\frac{1}{\hbar} \int_0^{\beta \hbar}d \tau \left( \frac{m\mathbf{\dot{\tilde{r}}}^2(\tau)}{2} + V[\tilde{\mathbf{r}}(\tau)] \right)}}}{\int_{\mathbf{r}(\beta \hbar)=\mathbf{r}(0)}{\mathcal{D}\mathbf{r}(\tau)e^{-\frac{1}{\hbar} \int_0^{\beta \hbar}d \tau \left( \frac{m\mathbf{\dot{r}}^2(\tau)}{2} + V[\mathbf{r}(\tau)] \right)}}}.
\label{end_to_end}
\end{equation}

The equations above allow us to express 
the NCP in terms of $\tilde{n}(\mathbf{x})$:


\begin{equation}
{J(y,\hat{\mathbf{q}})=\frac{1}{2\pi \hbar}\int{dx_{\parallel}\tilde{n}(x_{\parallel} \hat{\mathbf{q}})e^{\frac{i}{\hbar}x_{\parallel}y}},} 
\end{equation}

with $x_{\parallel}=\mathbf{x\cdot \hat{q}}$.

By making use of the primitive approximation \cite{Ceperley}:

\begin{equation}
{\tilde{n}(\mathbf{x})=e^{- \frac{m\mathbf{x}^2}{2 \beta \hbar^2}}e^{-\beta U(\mathbf{x})},}
\label{Trotter}
\end{equation}

the expressions for $U(x_{\parallel}\mathbf{\hat{q}})$ and the MF, $f(x_{\parallel}\mathbf{\hat{q}})$, become:

\begin{equation}
{U(x_{\parallel}\mathbf{\hat{q}})=- \frac{mx_{\parallel}^2}{2 \beta^2 \hbar^2}-\frac{1}{\beta}\ln \int{dy J(y,\mathbf{\hat{q}})e^{ix_{\parallel}y}}}
\label{potential}
\end{equation}

and 

\begin{equation}
f(x_{\parallel}\mathbf{\hat{q}})= - \frac{mx_{\parallel}}{\beta^2 \hbar^2}+\frac{1}{\beta}\frac{\int_0^{\infty}{y \sin (x_{\parallel}y)J(y,\mathbf{\hat{q}})dy}}{\int_0^{\infty}{dy \cos (x_{\parallel}y)J(y,\mathbf{\hat{q}})}},
\label{force}
\end{equation}

For finite temperature systems, Eq. \ref{Trotter}, which is valid for time step $\tau=\frac{\beta}{N} \rightarrow 0$, with $N$ the number of virtual 
replicas \cite{Ceperley}, is of particular relevance.

Indeed, the sequence of Feynman-Trotter approximations to the thermal Feynman path integral for a general non-relativistic system characterized by a smooth, 
single-minimum interaction potential converges pointwise to the quantum thermal propagator at every non-zero temperature, but in the zero-temperature limit, for 
high-order elements of the sequence, an abrupt \textquotedblleft collapse\textquotedblright \hspace{2mm} from the quantum to the classical ground-state takes place \cite{Kauffmann}.
In other words, for all \textit{finite} $N$-values the $T\rightarrow 0$ limit is unphysical. This situation can be mitigated by either increasing $N$ (which can prove computationally demanding) or 
implementing any alternative low-$T$ formulation to Feynman's original one, such as coherent-state path integral (CSPI) \cite{whitfield}.

In any case, Eq. \ref{potential} is an application of Feynman mapping of the quantum system onto a set of replicas obeying
to classical mechanics, where each particle in a string of replicas is referred to as a \textquotedblleft bead\textquotedblright, and adjacent beads
interact via a harmonic potential of frequency $\sim \sqrt{\frac{1}{\beta^2 \hbar^2}}$ \cite{Morrone2007}.

It is worth noticing here that the operational temperature mainly impacts the MF curve in terms of the slope of its local tangent in 
$x_{\parallel}=0$ $\textup{\AA}$, via the dominant negative
addend in Eq. \ref{force}.

\section{Mean Force in anisotropic harmonic potentials}

Previous DINS and simulation studies have described the momentum distribution of the hydrogen nuclei in water as a spherical average of a multivariate Gaussian according to \cite{Flammini2012}

\begin{equation}  
4\pi p^2 n(p)=\Big\langle \frac{\delta(p-|\mathbf{p}|)}{\sqrt{8 \pi^3}\sigma_x \sigma_y \sigma_z} \exp\left(-\frac{p_x^2}{2\sigma_x^2}-\frac{p_y^2}{2\sigma_y^2}-\frac{p_z^2}{2\sigma_z^2}\right)\Big\rangle,
\label{ndp}
\end{equation} 

where $\sigma_z$ is along the direction of the O-H bond, and $\sigma_x$ and $\sigma_y$ are in the plane perpendicular to the direction of the O-H bond. The set of parameters, $\sigma_{x,y,z}$, determines the anisotropy in the momentum distribution line shape, with:

\begin{equation}
\sigma_{i}^{2}=\frac{M\omega_{i}}{2\hbar}coth \left( \frac{\beta\hbar\omega_{i}}{2} \right),
\end{equation}

$\omega_i$ being an effective principal frequency \cite{Andreani,Andreani2013}.

The spherical average of $n(p)$ in Eq. \ref{ndp} is carried out over all possible molecular orientations, explicitly yielding:


\begin{equation}
n(p)=\frac{1}{4 \pi} \frac{1}{(2 \pi)^\frac{3}{2}} \frac{1}{\sigma_x \sigma_y \sigma_z} \int_0^{2 \pi}{d \phi}\int_0^{\pi}{[sin(\theta) e^{-\frac{1}{2}p^2 S(\theta,\phi)}]d\theta}. 
\label{num1}
\end{equation}

and the corresponding expression for the NCP is:

\begin{equation}
J(y)=\frac{\hbar}{2} \frac{1}{(2 \pi)^\frac{3}{2}} \frac{1}{\sigma_x \sigma_y \sigma_z}  \int_0^{2 \pi}{d \phi} \int_0^{\pi}{sin(\theta) \frac{1}{S(\theta,\phi)}e^{-\frac{1}{2}S(\theta,\phi)\hbar^2y^2}d\theta}
\label{num2}
\end{equation}

where

\begin{equation}
S(\theta,\phi)=\frac{sin^2(\theta)cos^2(\phi)}{\sigma_x^2}+\frac{sin^2(\theta)sin^2(\phi)}{\sigma_y^2}+\frac{cos^2(\theta)}{\sigma_z^2}. 
\end{equation}

Eqs. \ref{num1} and \ref{num2} can be evaluated numerically \cite{Andreani2016}.

For an isotropic system, the NCP is a univariate Gaussian, i.e., $J(y)$=$\frac{1}{\sqrt{2\pi}\sigma}e^{-\frac{y^2}{2 \sigma^2}}$ and the mean force in Eq. (\ref{force}) yields:

\begin{equation}
f(x_{\parallel}\mathbf{\hat{q}})= - \frac{mx_{\parallel}}{\beta^2 \hbar^2}+\frac{1}{\beta}\frac{{\frac{\sqrt{\pi}}{4}}(2 \sigma^2)^{\frac{3}{2}}x_{\parallel} e^{- \frac{\sigma^2}{2}x_{\parallel}^2}}{{\frac{\sqrt{\pi}}{2}} (2 \sigma^2)^{\frac{1}{2}} e^{- \frac{\sigma^2}{2}x_{\parallel}^2}}=\left( - \frac{m}{\beta^2 \hbar^2}+ \frac{\sigma^2}{\beta} \right) x_{\parallel}
\label{linear_curve}
\end{equation}

This expression for the mean force shows a linear dependence on the coordinate $x_{\parallel}$. Eq. (\ref{linear_curve}) is interpreted as Hooke's law governing the \textquotedblleft nanospring\textquotedblright \hspace{2mm} that connects two atoms along the direction of 
$x_{\parallel}$, with $k= \frac{m}{\beta^2 \hbar^2}- \frac{\sigma^2}{\beta}$ being the related elastic constant.
 In the anisotropic case, a three-dimensional harmonic potential would similarly produce such a linear behavior, when directional distributions are considered.

Therefore, deviations of $f(x_{\parallel}\mathbf{\hat{q}})$ from linearity provide evidence of underlying anharmonicity of the local potential. However, in experiments, only the spherically averaged NCP is accessible in liquids, amorphous solids, and
polycrystalline solids. The effect of the spherical average is to introduce deviations from linearity on $f(x_{\parallel}\mathbf{\hat{q}})$ \cite{Lin2010,Flammini2012}.

\begin{figure}[] 
\centering
\includegraphics[width=0.5\textwidth]{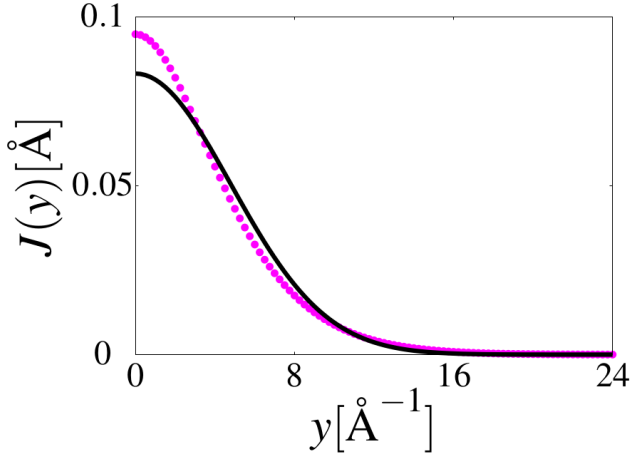}
\caption{
Hydrogen NCP $J(y)$ for $y>0$ at 100 K for: a) a system with a spherically averaged multivariate Gaussian $n(p)$, with 
$\sigma_x=2.0$ $\textup{\AA}^{-1}$, $\sigma_y=4.0$ $\textup{\AA}^{-1}$, $\sigma_z=7.0$ $\textup{\AA}^{-1}$, and $\overline{\sigma}_{aniso}=4.796$ 
$\textup{\AA}^{-1}$ (magenta points); b) a system with a univariate Gaussian $n(p)$, with $\sigma_{iso}=4.796$ $\textup{\AA}^{-1}$ (black line).}
\label{sph_average} 
\end{figure}

\begin{figure}[] 
\centering
\includegraphics[width=0.5\textwidth]{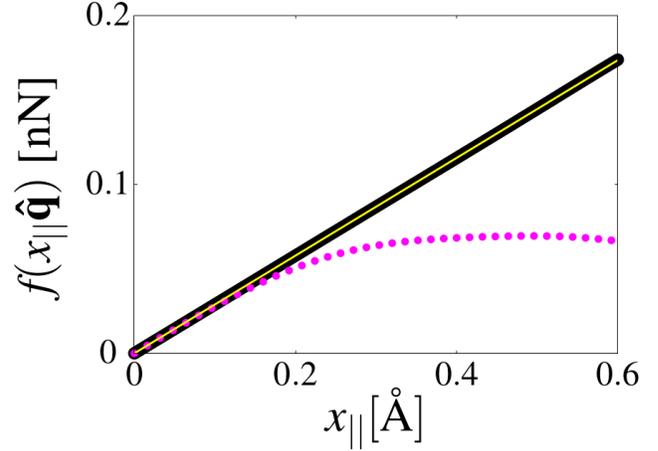}

\caption{
Hydrogen MF at $T$=100 K for: a) a system with a spherically averaged multivariate Gaussian $n(p)$, with 
$\sigma_x=2.0$ $\textup{\AA}^{-1}$, $\sigma_y=4.0$ $\textup{\AA}^{-1}$, $\sigma_z=7.0$ $\textup{\AA}^{-1}$, and $\overline{\sigma}_{aniso}=4.796$ 
$\textup{\AA}^{-1}$ 
(magenta points); b) a system with a univariate Gaussian $n(p)$ with $\sigma_{iso}=4.796$ $\textup{\AA}^{-1}$ (black line). The yellow line represents 
the analytical result from Eq. \ref{linear_curve}.}
\label{sph_average_2} 
\end{figure}

This is shown in Figs. \ref{sph_average} and \ref{sph_average_2}, where we report the calculated hydrogen NCP and MF, respectively, of a model system with a multivariate Gaussian momentum distribution characterized by a robust anisotropy (with average variance $\overline{\sigma}_{aniso}$ at $T$=100 K), together with the hydrogen NCP and MF of a model system with an isotropic momentum distribution at $T$=100 K, and $\sigma_{iso}=\overline{\sigma}_{aniso}$ .

Fig. \ref{sph_average_2} shows how, while the numerical evaluation performed on the 
isotropic system results in a linear MF overlapping the analytic
result from Eq. (\ref{linear_curve}), the MF for the anisotropic harmonic system only 
follows the linear trend for small values of $x_{\parallel}$, then shifting to a smoother
increase due to the spherical-averaging process. Here the deviation of the spherical force from linearity at
finite $x_{\parallel}$ results from the averaging process and is not a sign of anharmonicity.

We note that the effect of anisotropy is to introduce a concavity in the MF trend, but the slope of the resulting mean force is always lower than the slope of the corresponding isotropic reference system.

Thus, in the interpretation of the experimental Compton profiles, which result from the contribution of many particles, one must distinguish the case of an
anisotropic harmonic potential energy surface from that of an anharmonic potential energy surface. In view of the above, a practical means to identify anharmonicity in the experimental data is to compare the slope of the mean force with that of the corresponding isotropic model, which can be easily derived from the spherically averaged standard deviations of the neutron Compton profiles. 

\subsection{Evaluation of anharmonicity along the bond direction}

In this subsection, we show how one can fingerprint anharmonicity through the inspection of the mean force. 

Let us consider an anisotropic hydrogen-containing system at $T$=100 K. We describe the hydrogen $n(p)$ as a spherical average of an
anharmonic contribution along one direction, i.e., the $z$ axis, and two harmonic components along $x$ and $y$, respectively. 

\begin{figure}[] 
\centering
\includegraphics[width=0.4\textwidth]{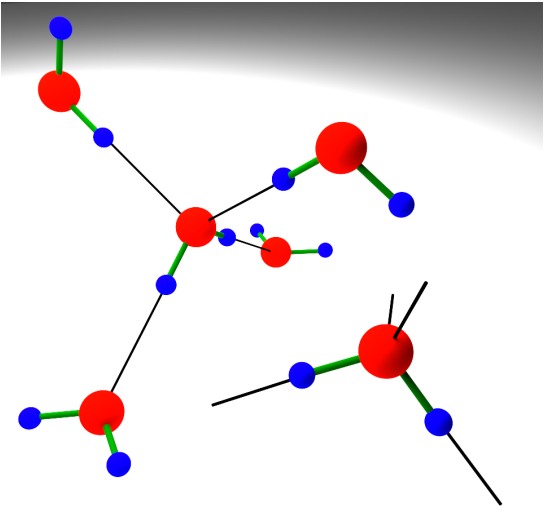}

\caption{
Schematics of a single water molecule (right) and its tetrahedral arrangement in low-temperature condensed phases (left).
O$-$H covalent bonds are in green, hydrogen bonds are in black.}
\label{water} 
\end{figure}

Let us suppose that a simple Morse potential is acting along the O$-$H covalent bond ($z$ axis), which, in
the typical tetrahedral arrangement of molecules in condensed water (Fig. \ref{water}), can be represented, to a first
approximation, as collinear with a hydrogen bond.

Of course, this type of modeling neglects the effects caused by both anharmonic coupling between covalent- and HB vibrations 
(in turn dependent on deviations from collinearity) and any departure from Morse-like 1D anharmonicity (in favor of a double-well potential), which can 
occur in stronger intermolecular hydrogen bonds \cite{Marechal2006}.

If the anharmonic contribution to the eigenfunction $\phi(\mathbf{p})$ of the ground state in the momentum space is represented as 
coming from a 
Morse-oscillator motion \cite{Morse1929,Dahl1988,Abramovitz1970}, then the spherical average of the momentum distribution yields:

\vspace{4mm}
$n(p)=\frac{1}{16 \pi^3} \frac{2^{\lambda_z-1}}{\sigma_x \sigma_y \alpha_z \hbar} \frac{2 \lambda_z-1}{\Gamma(2 \lambda_z)} \int_0^{2 \pi}{d \phi} \hspace{2mm} \int_0^{\pi}{d \theta \times}$

\begin{equation}
\times \left[ sin(\theta)  e^{- \frac{p^2}{2} \left( \frac{sin^2(\theta)cos^2(\phi)}{\sigma_x^2}+\frac{sin^2(\theta)sin^2(\phi)}{\sigma_y^2} \right)} \left| \Gamma \left( \lambda_z- \frac{1}{2}+i \frac{pcos(\theta)}{\alpha_z \hbar} \right) \right|^2 \right]
\end{equation}

\vspace{4mm}

and

\vspace{4mm}

$J(y)=2 \pi \hbar \int_{|\hbar y|}^{\infty}{p n(p) dp}=$

\vspace{4mm}

$=\frac{\hbar}{8 \pi^2} \frac{2^{\lambda_z-1}}{\sigma_x \sigma_y \alpha_z \hbar} \frac{2 \lambda_z-1}{\Gamma(2 \lambda_z)} \int_0^{2 \pi}{d \phi} \int_0^{\pi}{sin(\theta)d\theta} \hspace{2mm} \times$

\vspace{4mm}

\begin{equation}
\times  \int_{|\hbar y|}^{\infty} \left[ p e^{- \frac{p^2}{2} \left( \frac{sin^2(\theta)cos^2(\phi)}{\sigma_x^2}+\frac{sin^2(\theta)sin^2(\phi)}{\sigma_y^2} \right)} \hspace{1mm} \left| \Gamma \left( \lambda_z- \frac{1}{2}+i \frac{pcos(\theta)}{\alpha_z \hbar} \right) \right|^2 \right] \hspace{1mm} dp.
\end{equation}

We recall here that, since $D_z$ is the depth of the Morse potential minimum \cite{Morse1929}, $\alpha_z $ is its curvature, and $\omega_0=\sqrt{\frac{2D_z\alpha_z^2}{M}}$ is the harmonic fundamental frequency of the Morse oscillator, then $D_z \alpha_z^2$ can be derived from the analysis of the experimental line shapes through the following expression:

\begin{equation}
\sigma_z=\left[ \frac{M}{2 \hbar} \sqrt{\frac{2D_z\alpha_z^2}{M}} coth \left( \frac{\beta \hbar}{2} \sqrt{\frac{2D_z\alpha_z^2}{M}} \right) \right]^{\frac{1}{2}},
\end{equation}

where $\lambda_z=\frac{\sqrt{2MD_z}}{\alpha_z \hbar}$ is a dimensionless parameter, and either $\alpha_z$ or $D_z$ can be recovered from the literature.

\begin{figure}[] 
\centering
\includegraphics[width=0.5\textwidth]{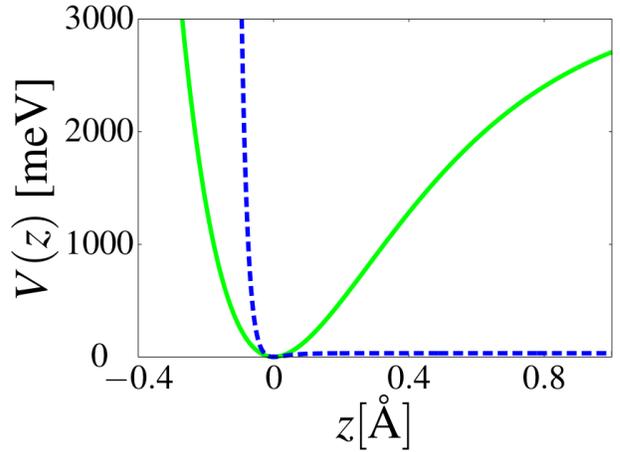}

\caption{
Shape of the Morse potential along the $z$ axis for $\sigma_z=7.0$ $\textup{\AA}^{-1}$, $\alpha_z=2.5$ $\textup{\AA}^{-1}$, and 
$D_z=3212$ meV (green line); and for $\sigma_z=7.0$ $\textup{\AA}^{-1}$, $\alpha_z=25$ $\textup{\AA}^{-1}$, and $D_z=32.12$ meV (blue dashed line).}
\label{Morse_pot} 
\end{figure}

The evaluation of the anharmonic contributions to the Morse potential shown above can be carried out by considering the anharmonic constant, $x_{anh}=\frac{\hbar \omega_0}{4D_z}=\frac{\hbar \alpha_z}{\sqrt{8D_z M}}=\frac{1}{2 \lambda_z}$. Using typical values of $D_z$ and $\lambda_z$ from the literature \cite{Lin2011,Lijima1994,McKenzie2014} we obtain $x_{anh}=0.032$. 
The resulting behavior of the mean force is expected to show small differences from the anisotropic harmonic case. More significant changes in the curvature of the mean force can be obtained by using $x_{anh}=3.2$, although this choice of the anharmonicity constant implies an almost flat-bottom potential. 
The two Morse potentials with $x_{anh}=0.032$ and $x_{anh}=3.2$ are shown in Fig. \ref{Morse_pot}.

Numerical calculation of $J(y)$ were carried out for the anisotropic harmonic model, and the anisotropic anharmonic model with the two anharmonic constants described above; the corresponding mean forces along the bond ($z$) direction, 
are shown in Fig. \ref{sph_average_3} below.

As anticipated above, considering a simple Morse potential along an O$-$H covalent bond collinear with a hydrogen bond amounts to 
neglecting, to a first approximation, a non-negligible portion of anharmonic behavior.

Indeed, hydrogen bonding normally introduce competing quantum effects on the vibrational motion of the hydrogen nucleus in the plane
of the 
water molecule, and perpendicular to it, due to its influence on the interplay between the O$-$H stretch and HB
bending \cite{Ceriotti2016}. These effects produce a $\textit{prominent}$
modification of the absorption band of the stretching mode of the X$-$H donor group, in terms of a red shift and spectral 
broadening \cite{Ceriotti2016,Hadzi1976}, which are a manifestation of the competition between the anharmonic quantum fluctuations of intramolecular 
covalent bond stretching and intermolecular hydrogen-bond bending \cite{Ceriotti2016}.

\begin{figure}[] 
\centering
\includegraphics[width=0.5\textwidth]{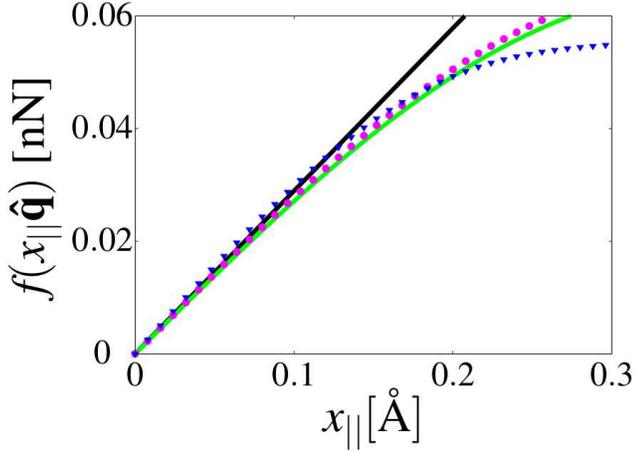}

\caption{
Hydrogen MF at $T$=100 K for: a) a system with a spherically averaged multivariate Gaussian $n(p)$, with 
$\sigma_x=2.0$ $\textup{\AA}^{-1}$, $\sigma_y=4.0$ $\textup{\AA}^{-1}$, $\sigma_z=7.0$ $\textup{\AA}^{-1}$, and $\overline{\sigma}_{aniso}=4.796$ $\textup{\AA}^{-1}$ 
(magenta points); b) same system as in a) with a spherically-averaged $n(p)$ with two harmonic components along the $x$ and $y$ directions and a small anharmonic component, $x_{anh}=0.032$, along the $z$ direction (green line); c) same system as in a) with a spherically-averaged $n(p)$ having two harmonic components along the $x$ and $y$ directions and a large anharmonic component ($x_{anh}=3.2$) along $z$ direction (blue triangles). The black 
line represents the analytical result along the bond direction from Eq. (\ref{linear_curve}).}
\label{sph_average_3} 
\end{figure}

In our first- approximation model, any large anharmonic effect may be modeled
by an increase of the anharmonic constant $x_{anh}$ along the bond direction, reflecting the enhancement of the non-parabolic character of the 
vibrational 
potential \cite{Barbatti2003}. This may lead, for example, to a very shallow model potential as seen in Fig. \ref{Morse_pot} (blue curve). 
Such a potential is clearly not adequate in describing the quantum state of the hydrogen nucleus in condensed water phases.

These findings suggest that an improved description of the anharmonicity due to the intermolecular hydrogen bonding has to take into
account the non-collinearity of the covalent-bond and hydrogen-bond 
directions, in the sense that each hydrogen atom should move under at least two non-collinear anharmonic potentials, one along 
the direction of the intramolecular 
O$-$H stretch and one along the intermolecular hydrogen bond \cite{Zhang2016}.
This model would imply that, differently from the previous cases, the motion along $x$-, $y$-, and $z$, and the 
corresponding contributions to the 
anisotropic $n$($p$) are correlated, due to the non-collinear potentials along the covalent and hydrogen bond directions.

Analogously, an upgrade of potential modeling to an asymmetric double-well \cite{Bakker2002} might be an interesting 
test.

Notwithstanding all of the above considerations, one must notice that the model presented in this paper is already suitable,
in the absence of biases due to the reduction of experimental data, to offer a practical, first-approximation tool to tell 
anharmonicity of the local potential apart from mere anisotropy. This is done by
assessment of the experimental MF possibly exceeding the slope of the isotropic linear reference.

\section{Application to experimental DINS data}

\subsection{Amorphous ices}

The determination of the hydrogen mean force from DINS data has been carried out in order to apply the models described in the previous sections to the interpretation of experimental data on amorphous and polycrystalline ice samples, with special attention to the identification of anharmonicities in the hydrogen nuclear effective potential. 

\begin{figure*}[] 
\centering

\includegraphics[width=0.95\textwidth]{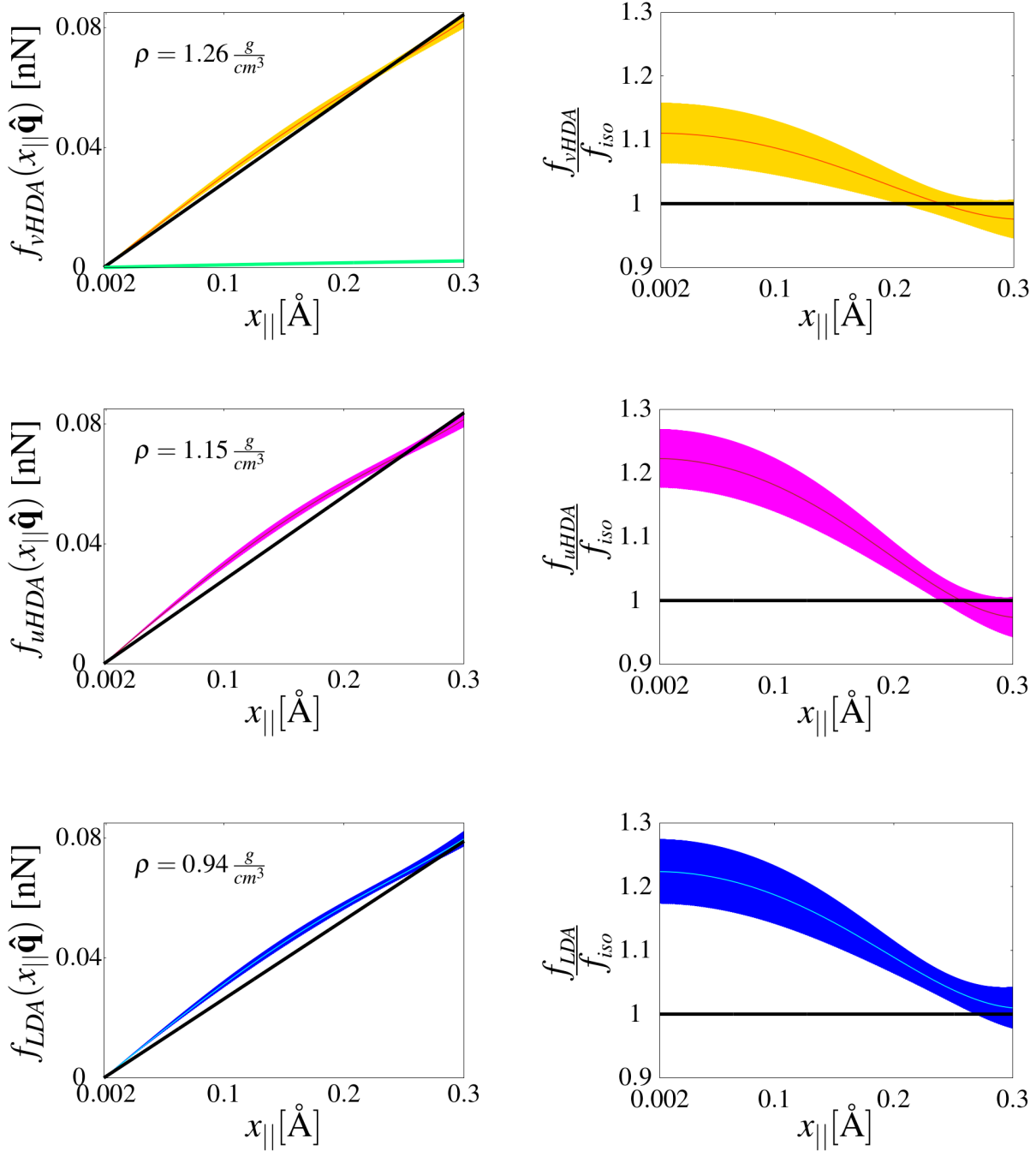}

\caption{Left panel: hydrogen MF along $\mathbf{\hat{q}}$ for vHDA (top, confidence region in yellow), uHDA (center, confidence region in magenta), and 
LDA (bottom, confidence region in blue), respectively, at $T$=80 K. The linear isotropic reference is obtained using a 1D Gaussian with 
$\sigma=\overline{\sigma}_{AI}$ (black). The turquoise line appearing in the plot for vHDA represents the mean force calculated using the detector-averaged resolution
profile as experimental data. AI densities signaled in the three plots are those reported in \cite{Parmentier2015}. Right panel: the same hydrogen MFs 
divided by the linear MF for the isotropic reference, in order to enhance MF curvature differences - with respect to the reference itself - as a function
of density. \newline
The $x_{\parallel}$-axis in each plot starts from 0.002 $\textup{\AA}$ since the point in the origin has zero error. This is caused by error bars
computed by error propagation from $J(y)$ alone.}
\label{AI_MF} 

\end{figure*}

The procedure adopted consists in the use of the single-detector NCPs, $F_l(y,q)= [J_{\rm IA} (y) + \Delta J_l(y, q)] \otimes R_l(y,q)$ 
(see equation \ref{defDY} , and the corresponding final-state-effect contributions obtained from the previous DINS experiments of References \cite{Flammini2012, Parmentier2015}). 
The single-detector NCPs were corrected for the final state effects and averaged over the set of individual detectors, yielding:
\begin{equation}
\hat{J}_R(y)= \langle  F_l(y,q)- \Delta J_l(y, q) \rangle_l.
\label{J_y_avg}
\end{equation}
Full details on the determination of $ F_l(y,q)$ and $\Delta J_l(y, q)$ can be found in References \cite{Flammini2012, Parmentier2015}.
The detector-averaged profile, $\hat{J}_R(y)$, contains the broadening due to the instrumental resolution. The latter is neglected to a first approximation in the present work, since its contribution is on the order of 15\% of the full width at half maximum of the neutron Compton profiles \cite{Senesi}. The hydrogen mean forces for the amorphous and polycrystalline samples are then determined by a numerical evaluation of Equation \ref{force} applied to $\hat{J}_R(y)$ for each sample.

At first we derive the mean force using DINS data from amorphous ices at $T$= 80 K and standard pressure \cite{Parmentier2015}, namely, 
very-high-density (vHDA), unannealed high-density (uHDA), and low-density (LDA) amorphous ices, respectively. The results, plotted in Fig. \ref{AI_MF}, 
are compared to model profiles from a univariate (1D) Gaussian momentum distribution with a standard deviation equal to the spherically averaged 
standard deviation, $\overline{\sigma}_{AI} $, found in Reference \cite{Parmentier2015}, that is, $\sigma=\overline{\sigma}_{AI}$ 
(Eq. \ref{linear_curve}).

Fig. \ref{AI_MF} shows that the hydrogen MFs in amorphous ices are characterized by a non-linear behavior, with slopes at intermediate $x$ 
exceeding those of the corresponding isotropic modeling with $\sigma=\overline{\sigma}_{AI}$. In view of the calculations and figures in Section 2, 
they provide evidence of anisotropic, as well as anharmonic, local environment for hydrogen nuclei. 
The right panel of Fig. \ref{AI_MF} clearly shows that, as density decreases (vHDA $\longrightarrow$ uHDA $\longrightarrow$ LDA) 
and within the experimental error, the
local radius of curvature of the MF becomes smaller, or the $x_{\parallel}$-range over which the MF is placed above its sample-specific 
isotropic reference becomes larger. This suggests that, in harmony with findings from inelastic and deep inelastic scattering measurements reported in Ref. \cite{Parmentier2015}, 
the hydrogen local environment in amorphous ices 
is characterized by an anharmonic character of the local potential, which goes beyond mere anisotropy, 
and decreases as density increases. 
This behavior is consistent with the structural differences found in the various forms of amorphous ices: indeed, 
all molecules in amorphous ices are hydrogen-bonded to four approximately tetrahedrally arranged neighbors (the
'Walrafen pentamer'). HDA holds an additional molecule (at a similar distance) not directly hydrogen bonded to the central molecule and located in between its first and second coordination shell; vHDA then holds two interstitial molecules \cite{Finney2002,Finney2002bis}.
We note that the structure of uHDA has recently been described as a \textquotedblleft derailed\textquotedblright \hspace{2mm} state along the ice I to ice IV pathway \cite{Shephard2017}.
 This picture is consistent with a longer average O$-$O distance
between hydrogen-bonded molecules for vHDA (2.85 $\textup{\AA}$)
than for uHDA (2.82 $\textup{\AA}$); an even shorter distance is found for LDA (2.77 $\textup{\AA}$) \cite{Guthrie2004}, in spite of a decrease in density and in the hydrogen mean kinetic energy \cite{Parmentier2015}.

In conclusion, as the hydrogen bonds weaken with increasing density, the effective potential becomes less and less anharmonic and its shape becomes similar to the shapes inferred from other spectroscopic techniques within the harmonic assumption \cite{Barbatti2003}.
This is consistent with recent findings on LDA and HDA from 2D IR spectroscopy \cite{Shalit2014}.

\subsection{Hexagonal ice I$\textit{h}$}

An early application of the MF formalism described in section
\ref{PMF} can be found in Ref. \cite{Flammini2012}, concerning DINS data from a polycrystalline sample of ice I$\textit{h}$ at
271 K and standard pressure.

\begin{figure}[]
\centering

\includegraphics[width=0.5\textwidth]{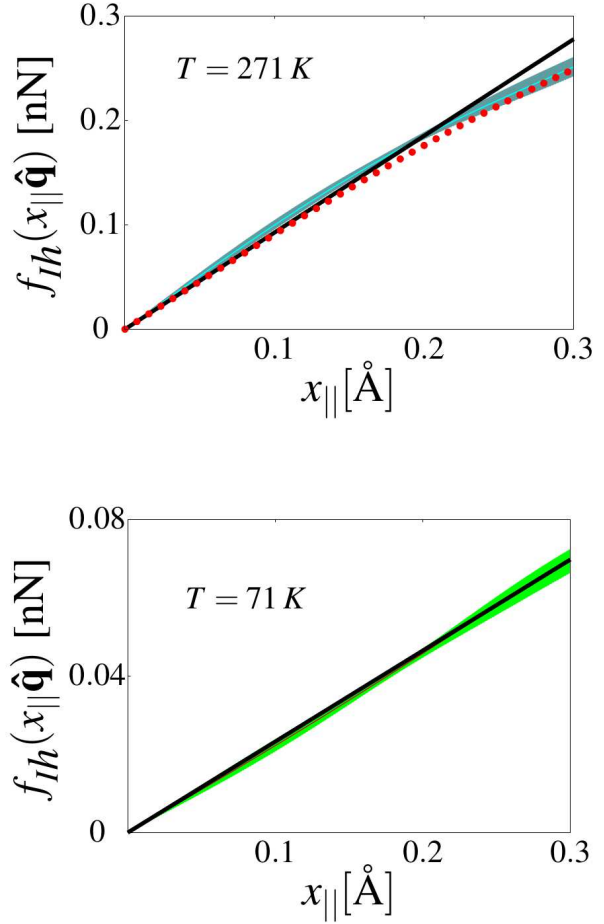}

\caption{
Top: Numerically evaluated hydrogen MF along $\mathbf{\hat{q}}$ for ice I$\textit{h}$ at 271 K and standard pressure (dark blue-green confidence region). 
In black we report the linear model obtained using a 1D Gaussian with $\sigma=\overline{\sigma}_{I\textit{h}}$. Red full points represent
the numerically evaluated MF of a spherically averaged harmonic anisotropic model system characterized by $\sigma_x=3.7$ $\textup{\AA}^{-1}$,
$\sigma_y=4.3$ $\textup{\AA}^{-1}$, and $\sigma_x=6.5$ $\textup{\AA}^{-1}$ \cite{Flammini2012}.
Bottom: Numerically evaluated hydrogen MF along $\mathbf{\hat{q}}$ for ice I$\textit{h}$ at 71 K and standard pressure (green confidence region). 
In black we report the linear model obtained using a 1D Gaussian with $\sigma=\overline{\sigma}_{I\textit{h}}$.}
\label{cryst_MF} 
\end{figure}

In order to assess whether the anharmonic behavior reported above 
might be resulting from systematic experimental contributions, we report below the determination of the hydrogen mean force
for a polycrystalline ice sample from a DINS measurement at 71 K and standard pressure \cite{Senesi2013}.

According to the literature \cite{Lin2011}, polycrystalline ice I$\textit{h}$ is expected to be a quasi-harmonic system,
especially at low temperature, with a small amount of anisotropy stemming from molecular orientations in the crystal.

 The MF derived from experimental DINS data for this system seems to confirm the above picture: as
reported in Fig. \ref{cryst_MF}, bottom panel, within experimental uncertainties the mean force in polycrystalline ice I$\textit{h}$ at 71 K does not show anharmonic behavior.

\section{Conclusions}

This paper presents a new procedure to obtain information on the hydrogen-nucleus energy surface in water by directly expressing the mean force function, $f$($x$), in terms of the neutron Compton profiles measured in DINS experiments, beyond what was introduced by Lin \textit{et al.} in \cite{Lin2010}. The new formalism is illustrated and applied to experimental DINS data in a variety of low-temperature
condensed phases of water.
The calculations on model systems allow to obtain a practical tool to identify anharmonicity in the hydrogen-nucleus effective potential, and to distinguish
the case of an anisotropic harmonic potential from that of an anharmonic potential, by simple inspection of the concavity and slope of the mean force.

By applying the above tools to the experimental data from DINS measurements, it is found that the shape of the mean force for amorphous and 
polycrystalline ice is primarily determined by the anisotropy of the underlying quasi-harmonic effective potential, and that data from amorphous ice 
show an additional curvature reflecting the more pronounced anharmonicity of the hydrogen-nucleus effective potential, with respect to that of ice 
I$\textit{h}$.

The present work joins the stream of efforts to better understand the relation between the experimental momentum distribution and the 
atomic/molecular environment: a challenging task, though crucial to the study of condensed matter and worthy of development.

We plan to further refine the proposed approach to better formalize the role of molecular hydrogen bonding with the aim of quantitative determinations of the anharmonicities of atomic local potentials.

\section*{Acknowledgments}
This work was supported within the CNR-STFC Agreement 2014-2020 (No. 3420 2014-2020) concerning collaboration in scientific research at ISIS Spallation Neutron Source.


\end{document}